\PassOptionsToPackage{letterspace=-20,tracking=true,final}{microtype}
\documentclass[sigconf]{acmart}

\AtBeginDocument{%
  \providecommand\BibTeX{{%
    \normalfont B\kern-0.5em{\scshape i\kern-0.25em b}\kern-0.8em\TeX}} }
\setcopyright{none} 
\copyrightyear{2025}
\acmYear{2025}
\acmDOI{XX.XXXX}
\acmConference[arXiv]{February, 2025}{February,2025}{Barcelona, Spain}
\acmISBN{978-1-4503-XXXX-X/18/06}
\usepackage{mathtools}

\usepackage{algorithm}
\usepackage{algpseudocode}

\usepackage{graphicx}
\usepackage{textcomp}
\usepackage[normalem]{ulem}
\usepackage{color}
\usepackage{xcolor}
\usepackage{caption}
\usepackage{subcaption}

\usepackage{array}
\usepackage{booktabs}
\usepackage{colortbl}
\usepackage{multirow}

\def\BibTeX{{\rm B\kern-.05em{\sc i\kern-.025em b}\kern-.08em
    T\kern-.1667em\lower.7ex\hbox{E}\kern-.125emX}}

\title{Hamun: An Approximate Computation Method to Prolong the Lifespan of ReRAM-Based Accelerators}

\author{Mohammad Sabri, Marc Riera, Antonio González}
\email{mohammad.sabri@upc.edu, marc.riera.villanueva@upc.edu, antonio@ac.upc.edu}
\affiliation{
  \institution{Universitat Polit\`{e}cnica de Catalunya (UPC)}
  \city{Barcelona}
  \country{Spain}
}
\renewcommand{\shortauthors}{Sabri et al.}

\usepackage{microtype}

\begin{document}

\begin{abstract}
ReRAM-based accelerators exhibit enormous potential to increase computational efficiency for DNN inference tasks, delivering significant performance and energy savings over traditional platforms. By incorporating adaptive scheduling, these accelerators dynamically adjust to DNN requirements, optimizing allocation of constrained hardware resources. However, ReRAM cells have limited endurance cycles due to wear-out from multiple updates for each inference execution, which shortens the lifespan of ReRAM-based accelerators and presents a practical challenge in positioning them as alternatives to conventional platforms like TPUs. Addressing these endurance limitations is essential for making ReRAM-based solutions viable for long-term, high-performance DNN inference.


To address the lifespan limitations of ReRAM-based accelerators, we introduce \textit{Hamun}, an approximate computing method designed to extend the lifespan of ReRAM-based accelerators through a range of optimizations. Hamun incorporates a novel mechanism that detects faulty cell due to wear-out and retires them, avoiding in this way their otherwise adverse impact on DNN accuracy. Moreover, Hamun extends the lifespan of ReRAM-based accelerators by adapting wear-leveling techniques across various abstraction levels of the accelerator and implementing a batch execution scheme to maximize ReRAM cell usage for multiple inferences. Additionally, Hamun introduces a new approximation method that leverages the fault tolerance characteristics of DNNs to delay the retirement of worn-out cells, reducing the performance penalty of retired  and further extending the accelerator’s lifespan. On average, evaluated on a set of popular DNNs, Hamun demonstrates an improvement in lifespan of $13.2\times$ over a state-of-the-art baseline. The main contributors to this improvement are the fault handling and batch execution schemes, which provide $4.6\times$ and $2.6\times$ lifespan improvements respectively.
\end{abstract}

\maketitle

\textls{
\vspace{-.5cm}

\section{Introduction}\label{s:intro}
Processing Using Memory (PUM) offers a promising solution to reduce the high computational and energy demands of Deep Neural Network (DNN) inference. Several Non-Volatile Memory (NVM) technologies, such as Phase-Change Memory (PCM)~\cite{PCM}, Spin-Transfer Torque Magnetic RAM (STT-MRAM)~\cite{STT}, and Resistive RAM (ReRAM)~\cite{ResiRCA}, can effectively implement the critical dot-product operations for DNNs in the memory arrays. Among these, ReRAM stands out due to its lower read latency and higher density, making it an ideal choice for DNN accelerators. As a result, there is increasing interest in ReRAM-based PUM designs for their dense and efficient computation capabilities~\cite{PRIME, sparse_ReRAM, CASCADE}.

While ReRAM offers many advantages, its high write latency and energy consumption pose significant challenges in designing efficient DNN accelerators. To address these limitations, prior research has proposed executing multiple DNN inferences in a pipelined fashion to reduce the write overheads of ReRAM~\cite{RAPIDNN, FORMS, ISAAC, TIMELY, AtomLayer, NVMExplorer, RAELLA, Neurosim_mapping, PUMA}. These accelerators are typically designed with the assumption that there are sufficient ReRAM cells to store all network weights, meaning that DNN weights are written only once at the start of the first inference. Although this method minimizes ReRAM's write inefficiencies, it introduces a new challenge that could hinder the future viability of ReRAM-based accelerators as a leading computational platform.

As DNNs continue to grow in complexity, scaling an accelerator's resources to accommodate the largest models is neither area- nor power-efficient. Additionally, this approach lacks \textit{adaptability}, as the accelerators used at present may be unable to handle future, larger networks. To address this problem, some recent proposals~\cite{ARAS, MNEMOSENE, PipeLayer} introduced adaptive, ReRAM-based accelerators for DNN inference that efficiently write DNN weights into a limited-size accelerator at runtime. Unlike previous accelerators, which assume all DNN weights are preloaded, these solutions employ an scheduler to manage tasks such as data transfers, weight updates, and dot-product computations.


As an example, ARAS~\cite{ARAS} introduces an offline scheduler and a novel execution scheme that allocates available ReRAM cells and perform weight updates of upcoming layers concurrently with the computations in the current layer. Computations are performed sequentially, layer by layer, due to the data dependencies across layers. This approach allows the accelerator to networks of unlimited size with limited resources, and maximizes resource utilization. While ARAS, and other previous proposals, addresses the \textit{adaptability} challenge and introduces new optimizations to reduce the energy and latency costs of ReRAM writes, frequent ReRAM cell updates to allocate resources for upcoming layers remain an important problem due to the limited write endurance of ReRAM cells~\cite{ReRAM_Challenges, ReRAM_characterizing, Aging}.

Current commercial ReRAM storage chips~\cite{Weebit, crossbar} exhibit much lower endurance compared to other memory technologies. This reveals a significant gap in the development of ReRAM cells optimized for PUM applications. Consequently, the lifespan of ReRAM-based accelerators is compromised due to the frequent cell updates required for DNN inference. In this paper, we propose \textit{Hamun\footnotemark}, an approximate computation method to prolong the lifespan of ReRAM-based accelerators. The key innovation is a scheduler that manages weight writings across ReRAM cells, and continues to work even when some cells completely wear out by leveraging the intrinsic fault tolerance of DNN inference. The scheduler includes a mechanism that retires faulty cells once their impact on accuracy exceeds the error tolerance of the DNN.

\footnotetext{A lake that requires special attention to ensure their long-term sustainability.}

Hamun scheduler, similar to ARAS, performs static binding and scheduling tasks for a given DNN, orchestrating the instructions that the accelerator will execute during each inference. When a cell wears out during the weights update for a particular layer, the accelerator sends the location of the faulty cell to the host. The host identifies which network layers are affected by the faulty cell for the current binding configuration and estimates the potential accuracy loss caused by the new fault. If the accuracy impact remains within the DNN’s fault tolerance limits, the accelerator continues with the existing binding and scheduling configuration. Otherwise, all faulty cells are retired, and the scheduler generates a new binding and scheduling configuration that excludes them.

As previously mentioned, this paper introduces a procedure to estimate the accuracy impact of faulty cells. This procedure computes a strict threshold for the number of tolerable faults per each network layer. If the number of faults in a layer exceeds this threshold, the scheduler sets a new binding configuration that excludes the faulty cells. This method ensures that the accuracy loss does not exceed a user-specified limit. The threshold for each layer is determined through a static process that incrementally impose random faults into the network model until the accuracy loss surpasses the defined limit.


Hamun scheduler incorporates various optimizations to reduce the frequency of ReRAM cell updates during each inference and reuse the weights for multiple inferences. In particular, it evenly distributes writes across all cells, preventing premature wear-out of specific cells. Another optimization is batch execution, where multiple inferences are processed simultaneously. In this way, weights of a layer are reused across all inferences within a batch before updating the weights for the next layers. Partial results are stored in an on-chip SRAM buffer, and the buffer size determines the maximum number of inferences that can be processed in a batch.


In summary, this work aims to extend the lifespan of ReRAM accelerators by using an approximation technique that delays the retirement of wear-out cells if their impact on accuracy is negligible. The main contributions are:

\begin{itemize}

\item We introduce a scheduler that performs binding and scheduling tasks to maximize the lifespan of ReRAM-based accelerators. The scheduler incorporates various optimizations to reuse weights both within and across inferences, and it ensures even utilization of all ReRAM resources to prevent premature wear-out of specific cells. In addition, the scheduler includes a mechanism that retires faulty cells once their impact on accuracy exceeds the DNN’s error tolerance.


\item The scheduler estimates how faulty cells might affect overall accuracy. When the estimated impact is within an acceptable range, as defined by the user, it delays the retirement of these cells to help extend their useful lifespan.

\item We evaluate the proposed scheme, called \textit{Hamun}, on three representative DNNs. On average, \textit{Hamun}, with all its optimizations, extends the lifespan of the accelerator by $13.2\times$. Notably, the improvements of $4.6\times$ and $2.6\times$ are directly attributed to the fault-handling and batching schemes, respectively.

\end{itemize}



\section{Background and Related work}\label{s:Background}
In this section, we provide some background on ReRAM writing schemes and techniques used to prolong the lifespan of ReRAM cells. Additionally, we describe the in-situ matrix transposition operation. Finally, we discuss the architecture of popular Transformer-based neural networks.

\subsection{ReRAM Cell and Crossbar Architecture}\label{subs:Writing}
ReRAM cells are two-terminal devices (see Figure~\ref{fig:CrossBars}(b)) capable of storing values through their multilevel conductance states. In DNN accelerators, ReRAM cells store network weights. The process of writing these weights (i.e., weight updates) involves transitions between different conductance states, typically triggered by electrical inputs. The specific mechanism driving these transitions varies depending on the type of ReRAM insulator used~\cite{insulator, insulator2, Park_2013, Woo_2016}. Generally, the conductance of ReRAM cells is increased or decreased through positive and negative programming voltage pulses, corresponding to weight increases and decreases, respectively. Ideally, ReRAM cells would exhibit a linear weight update response to identical programming voltage pulses. However, real-world devices, as documented in the literature~\cite{trade-offs_MLC,Gao_2015}, deviate from this ideal behavior and exhibit "non-ideal" properties. In addition, process variations impact the ReRAM cells specifications, both within the same device and across different devices.

\begin{figure}[t!]
    \centering
    \includegraphics[width=0.7\columnwidth]{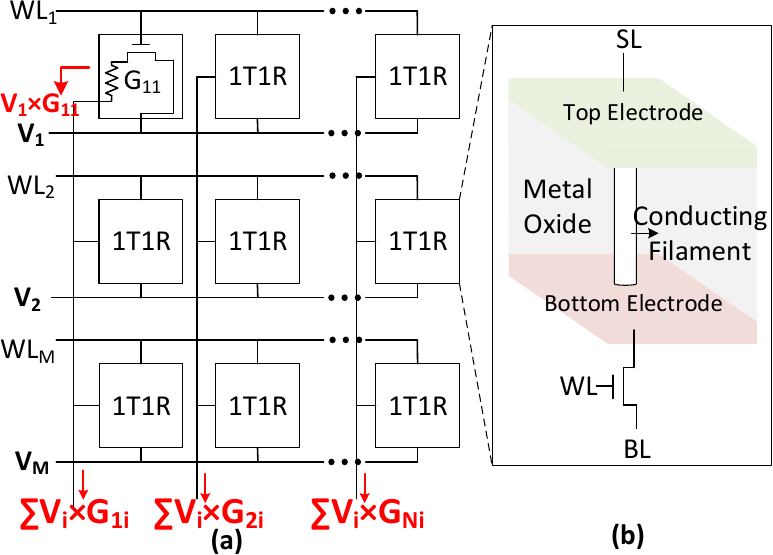}
    \caption{Crossbar architecture and analog operation with ReRAM cells (a), ReRAM cell device (b).}
    \label{fig:CrossBars}
\end{figure}

To handle the impact of non-ideality on multi-level cell (MLC) and enhance tolerance to process variations, the Program and Verify (P\&V) method has been proposed~\cite{Accurate_Program_Verify,ISPVA}. P\&V involves applying narrow voltage pulses with progressively increasing magnitude, combined with read pulses to verify whether the cell has reached the desired state, as shown in Figure~\ref{fig:P&V}. This approach not only accounts for process variation but also helps identify worn-out cells, as those that are stuck at a certain value will not respond to further programming pulses. As illustrated in Figure~\ref{fig:P&V}, the programming pulses have a higher amplitude and longer duration than the read pulses, meaning the additional read pulses introduce minimal overhead. According to \cite{Assist_Techniques}, the P\&V method increases energy consumption and write latency by only 5\% and 6\%, respectively, compared to an "ideal" cell.

The most compact and straightforward structure for storing a DNN weight matrix in ReRAM cells is the crossbar array configuration, where each 1T1R cell is positioned at the intersection points. Crossbars provide the benefit of high integration density~\cite{NeuroSim_writing} and serve a dual function: storing weights and performing dot-product operations between inputs and weights. As illustrated in Figure~\ref{fig:CrossBars}(a), the input vector is encoded into read voltage signals, allowing for the parallel execution of matrix-vector multiplication using the weights stored in the crossbar. The resulting weighted sums are captured at the end of each column as analog currents, which are then post-processed by additional peripherals to convert them into digital values.

\begin{figure}[t!]
    \centering
    \includegraphics[width=0.70\columnwidth]{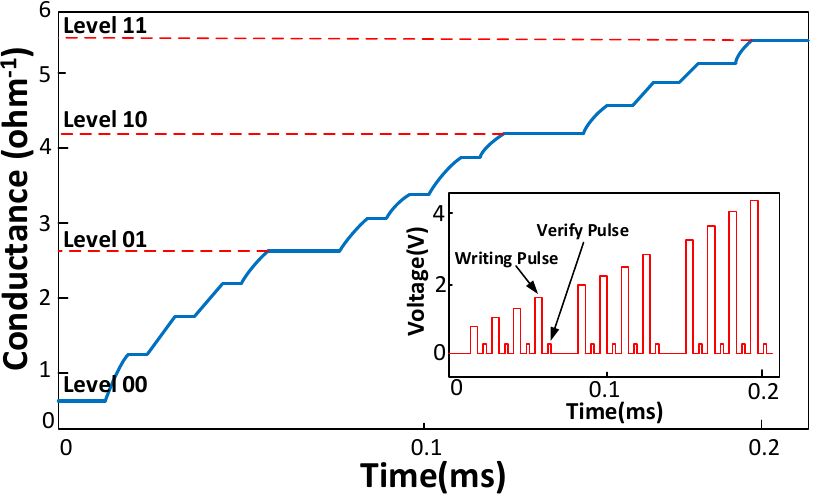}
    \caption{Program and verify (P\&V).}
    \label{fig:P&V}
\end{figure}

To update weights in ReRAM crossbars, NeuroSim~\cite{NeuroSim_writing} introduce a row-by-row writing which is typically employed, as shown in Figure~\ref{fig:WeightUpdating}. The weight increase and decrease operations require distinct programming voltage polarities, so the weight update process is divided into two steps, each with its specific voltage polarity for every row. In each step, the SLs (Source Lines) provide voltage pulses or constant voltages (if no update) to modify each selected cell, while the BL (Bit Line) provides the required polarity for that particular step. It is important to note that the update of a each cell's value follows the P\&V scheme, which is achieved by applying specific pulses through the associated SL and BL, and the number and amplitude of pulses required are determined by the current cell value and the desired target value. Consequently, in a crossbar array, each SL is equipped with its independent driver to deliver distinct pulses.

\begin{figure}[t!]
    \centering
    \includegraphics[width=0.8\columnwidth]{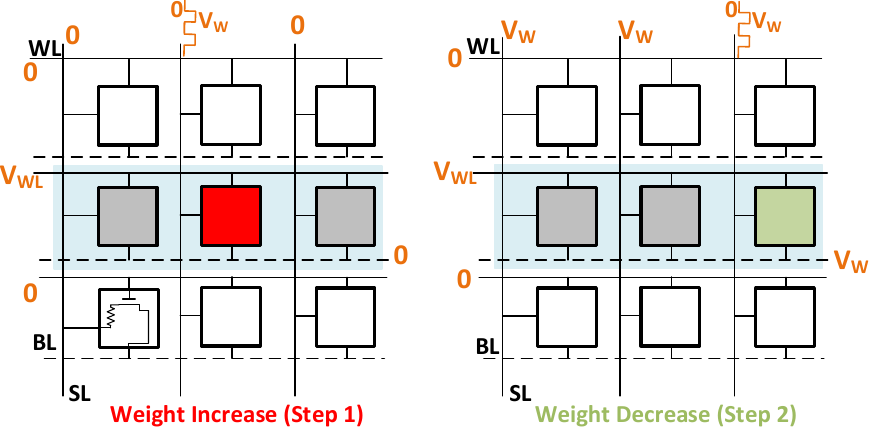}
    \vskip -0.10in
    \caption{Example of a Row-by-Row writing scheme. The gray cells keep their values in each steps, while the red and green cell values are modified.}
    \label{fig:WeightUpdating}
    \vskip -0.15in
\end{figure}

\subsection{ReRAM Lifespan Improvement Techniques}\label{subs:Lifespan improvements}
NVMs, particularly ReRAMs, suffer from limited write endurance, causing cells to become stuck at a fixed value (permanent failure) when the number of write cycles exceeds the cell's endurance. For ReRAM cells, this limit typically occurs after $10^6$ to $10^9$ write cycles on average~\cite{Weebit, crossbar, Realizing}. To extend memory lifespan, two primary solutions are wear-leveling (WL) and fault tolerance techniques~\cite{FLOWER_and_FaME, WoLFRaM}.

\textbf{Wear-leveling techniques.} These techniques aim to distribute writes, which are often concentrated on a small subset of cells, more evenly across the entire memory~\cite{WoLFRaM,Realizing,ARS}. As a result, uniform wear-leveling (UWL) represents the upper bound of wear-leveling methods. Wear-leveling techniques are applied at various levels of the memory hierarchy, whether using ReRAM for storage or in DNN accelerators. For instance, row-level wear leveling extends the lifespan of each memory page (ReRAM cells for storage) or processing element (ReRAM cells for computation) by balancing writes across rows~\cite{ARS, SAWL,Bloom_filter}. Similarly, column-level wear leveling balances write operations across cells in the same row by rotating the order of writes across columns at different granularities~\cite{WELCOMF,ILF}. In \cite{ReNEW,On_Endurance}, row-level and column-level wear leveling techniques are adapted for ReRAM-based accelerators. However, these approaches still suffer from a lack of \textit{adaptability}, as they assume the accelerator has sufficient capacity to store all the required data.



\textbf{Fault tolerance techniques.} Extensive research has been conducted to address stuck-at faults caused by limited endurance in NVMs~\cite{FLOWER_and_FaME,FREE-p,Data_Block,ECP}. For instance, RETROFIT~\cite{Realizing} employs Error Correction Codes (ECC) to correct faults in worn-out cells and uses ECC to identify "weak" rows, which are then leveraged by row-level wear leveling techniques. However, to the best of our knowledge, no research has explored ReRAM-based accelerators for DNN inference that leverage the inherent fault tolerance of DNNs to mitigate the negative impact of worn-out cells.


\subsection{In-Situ Matrix Transposition}\label{subs:Transposition}
Matrix transposition is a fundamental operation in many DNNs, including architectures like Transformers and GPT models. Performing matrix transposition with standard methods can be resource-intensive, as these approaches often require extra memory to hold both the original and transposed matrices. This additional memory requirement can be costly, particularly in large-scale DNNs where efficient memory management is critical to sustaining performance and minimizing latency. 

In-place matrix transposition has been widely studied, with early work dating back to 1967 and 1970~\cite{Algorithm_380,Algorithm_302}. However, in 1977, Cate and Twigg introduced Algorithm 513~\cite{Algorithm_513}, which proposed an efficient approach for matrix transposition directly within memory, avoiding the need for additional memory allocation. This algorithm focuses on optimizing the transposition process by rearranging matrix elements stored in a one-dimensional array—often referred to as a flattened matrix. The primary advantage of this method is that it handles transposition in-situ (without extra storage), which is particularly beneficial for memory-constrained environments or when working with large matrices. For example, consider $2\times4$ matrix and its transposed:

\begin{equation}
  M = 
  \begin{bmatrix}
    101 & 102 & 103 & 104 \\
    201 & 202 & 203 & 204 \\
  \end{bmatrix}
  , M^{T} =
  \begin{bmatrix}
    101 & 201 \\
    102 & 202 \\
    103 & 203\\
    104 & 204\\
  \end{bmatrix} 
\end{equation}

Figure~\ref{fig:Transposition} shows the flattened representation of matrix $M$. The transposed matrix can be obtained by applying specific permutations to the indices. In this example, the permutation is divided into four distinct cycles: $\{0\}$, $\{1,2,4\}$, $\{3,6,5\}$, and $\{7\}$. Each cycle represents a set of indices that need to be swapped. For instance, the cycle $\{1,2,4\}$ indicates that the first and second indices should be swapped, followed by swapping the first and fourth indices, ensuring that the correct transposition is achieved. Cycles that contain only one index, like \{0\} and \{7\}, indicate fixed points, where the index maps to the same position in both the original and transposed matrix, as described in~\cite{Algorithm_513}.

Equation~\ref{eq:Transposition} defines a straightforward formula that identifies the position of an element $\alpha$ in a flattened matrix of size $N \times M$ and determines where it maps in the transposed matrix. This mapping is used when performing an in-place matrix transposition, ensuring that the element at a given index in the original flattened matrix is correctly relocated to its corresponding index in the transposed version.

\begin{equation}
    P(\alpha) =\left\{
      \begin{array}{@{}ll@{}}
        MN-1, & \text{if}\ \alpha=MN-1 \\
        N\alpha \ mod \ (MN-1), & \text{otherwise}
      \end{array}\right. 
      \label{eq:Transposition}
\end{equation}

\begin{figure}[t!]
    \centering
    \includegraphics[width=0.70\columnwidth]{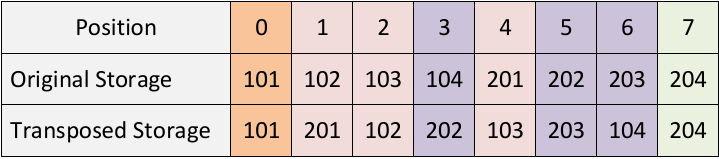}
    \vskip -0.10in
    \caption{Flattened representation of matrix $M$ and its transposed form. Each colored set represents a cycle in Algorithm 513, indicating which elements need to be swapped in order to generate the transposed matrix.}
    \label{fig:Transposition}
    \vskip -0.10in
\end{figure}


\subsection{Transformer DNNs}\label{subs:Transformers}
Transformers are built from several stacked encoders and decoders that process input sequences through self-attention and feed-forward networks (FFNs) like Fig~\ref{fig:Transformer}. In the encoder, each token in the input sequence is transformed into query, key and value vectors using fully-connected (FC) layers. Subsequently, by combining these vectors across all tokens, the query ($Q$), key ($K$), and value ($V$) matrices are produced for the entire input sequence. To compute the self-attention scores, the query matrix $Q$ is multiplied by the transposed key matrix $K$ and then scaled, to reveal relationships between tokens. After softmax normalization and dropout, these scores form the self-attention matrix, which is multiplied by $V$ to generate each head output. Multiple attention heads run in parallel to capture diverse patterns, and their outputs are combined and processed through an FC layer. Afterwards, a residual connection and normalization layer stabilize the output values. The processed attention output then passes through an FFN, which generates the encoder's final output.

The output of each encoder or decoder block in a Transformer model can either be passed as input to subsequent layers or directed to a task-specific output layer, such as a classification layer. Decoder blocks follow a structure and flow similar to encoder blocks but with some key differences. Unlike encoder blocks, which handle sequences of tokens, decoder blocks, used in generation tasks, produce tokens one-by-one, feeding each generated token back as input for the next token's generation.

\begin{figure}[t!]
    \centering
    \includegraphics[width=0.80\columnwidth]{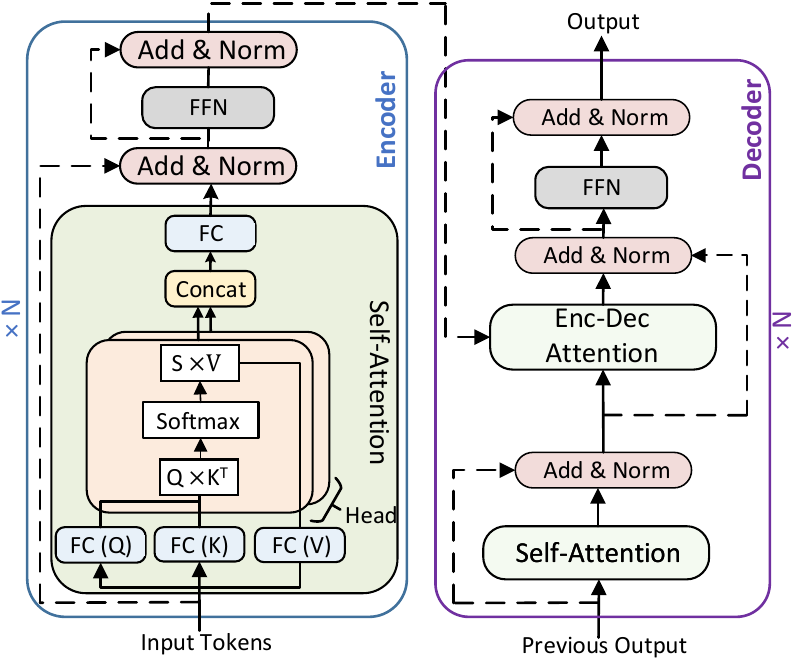}
    \vskip -0.10in
    \caption{Diagram of the Transformer architecture, featuring $N$ sequential encoder blocks followed by $N$ sequential decoder blocks.}
    \label{fig:Transformer}
    \vskip -0.15in
\end{figure}

\section{Hamun Accelerator}\label{s:Accelerator}
In this section, we describe the hardware architecture of our proposed ReRAM-based accelerator, which is organized into multiple hierarchical levels. We then explain the execution flow, detailing the weight-writing and computation procedures.

\subsection{Architecture}\label{subs:Architecture}
Most prior works (e.g.,~\cite{RAELLA, ISAAC}) design accelerators under the assumption that all DNN weights are pre-stored in ReRAM crossbars. This approach demands substantial resources, including large numbers of ReRAM crossbars and on-chip SRAM buffers, making it inefficient for large networks. In contrast, proposals like ARAS~\cite{ARAS} operate with limited resources by adopting a layer-by-layer execution, where the computation of one layer overlaps with the writing of weights for subsequent layer(s). However, the frequent updates to ReRAM cells for executing DNN layers reduce the lifespan of the accelerator due to ReRAM's limited endurance cycles. This is one of the key challenges that Hamun addresses by proposing an approximate computing scheme along with some wear-leveling optimizations.

Figure~\ref{fig:Top_view} presents a top-down view of the Hamun accelerator architecture. Figure~\ref{fig:Top_view}(a) shows a high-level schematic of the chip architecture. A single chip comprises several key components: an External IO Interface (Ext-IO), multiple Processing Elements (PEs), a Special Function Unit (SFU), Accumulation Units (ACC), and a Global Buffer (Gbuffer). The Ext-IO facilitates communication with Main Memory (MM), loading network weights and inputs, and storing the final outputs. The ACC units aggregate partial results from neural operations for a given layer, which may be generated across multiple PEs when a layer spans several PEs due to its size. The SFU handles transitional operations such as pooling, non-linear activations (e.g., sigmoid or ReLU), and normalization, ensuring support for the full range of computations required for state-of-the-art DNNs. The Global Buffer (Gbuffer) acts as a storage unit for intermediate activations produced during the execution of each layer.

\begin{figure}[t!]
    \centering
    \includegraphics[width=0.95\columnwidth]{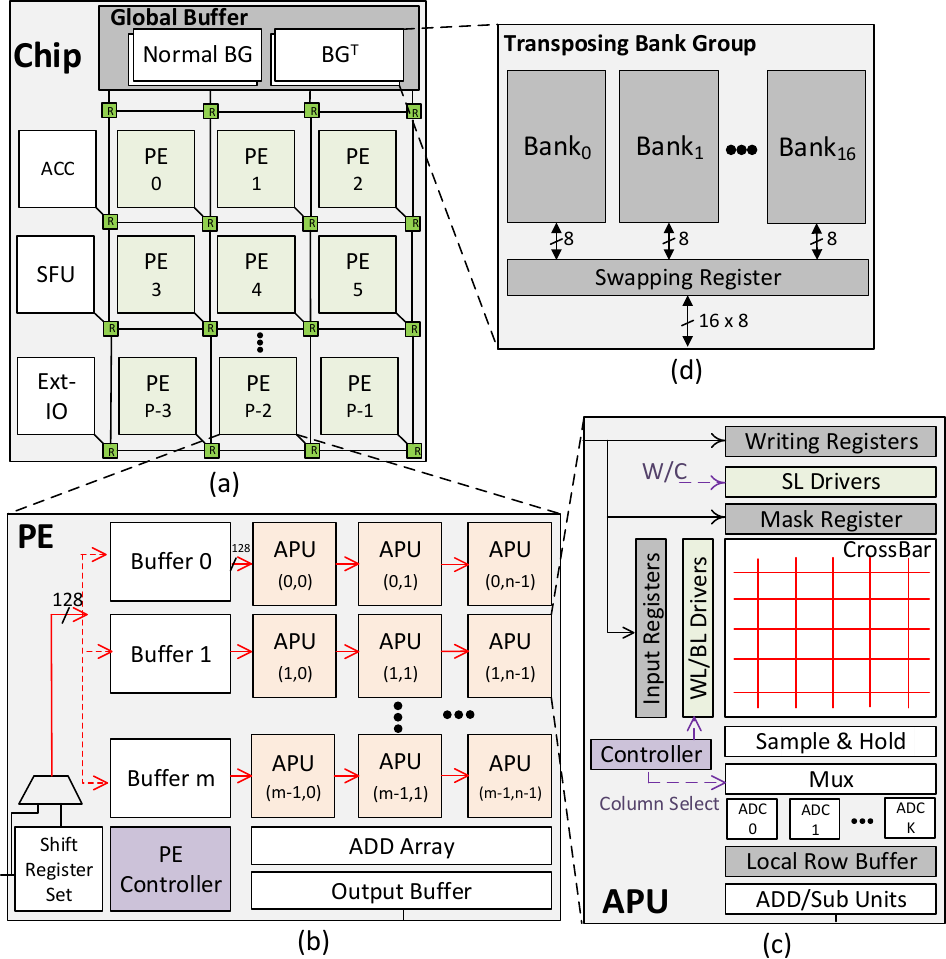}
    \vskip -0.10in
    \caption{Architecture of the Hamun accelerator including the organization of: (a) Chip, (b) Processing Element (PE), (c) Analog Processing Unit (APU), and (d) Transposing Banks.}
    \vskip -0.15in
    \label{fig:Top_view}
\end{figure}

Figure~\ref{fig:Top_view}(b) illustrates the structure of a Processing Element (PE), which consists of $m \times n$ Analog Processing Units (APUs), $m$ buffers to store either input activations or weights, an output buffer for storing partial sums, $n$ accumulation modules to sum the partial outputs from the APUs in each column, shift registers to serialize activations, and a multiplexer that switches between the weight-writing and computation phases. Figure~\ref{fig:Top_view}(c) displays the main components of an Analog Processing Unit (APU), which includes a ReRAM-based crossbar array of 1T1R cells used to store synaptic weights. Each APU also includes an input register, which is used to store activations, a write register to store weights, a Mask register to disable an entire crossbar column in the writing/computing procedure, a WL/BL (Wordline/Bitline) driver to control the wordlines and bitlines, and an SL (Sourceline) driver to generate required voltage pulses in the sourcelines. Moreover, each APU has an analog multiplexer, a shared pool of Analog-to-Digital Converters (ADCs), and functional units for accumulating and shifting the partial results from different bitlines across iterations. Further details on the analog dot-product operation procedure using ReRAM crossbars, as well as the ReRAM cell writing process, can be found in Section~\ref{subs:Execution Dataflow}.

The Gbuffer is divided into different bank groups. While one bank group operates in a standard mode, the other bank group follows a distinct writing procedure that results in a transposed matrix format when reading data. As discussed in Section~\ref{subs:Transformers}, in the attention block, calculating the score matrix requires performing matrix multiplication between the query matrix and the transposed key matrix. This scheme facilitates the transposition of the key matrix in-situ, without introducing any additional latency overhead.

Figure~\ref{fig:Example_In_Transposition} provides a simplified example illustrating our proposed in-situ transposition scheme. In (a) we present an original matrix that needs transposing. The matrix elements are generated sequentially in a row-by-row manner and transferred via the NoC to storage. For this example, the NoC transmits only two elements per transaction. These are then stored in two memory banks for each transfer. Figure~\ref{fig:Example_In_Transposition}(b) depicts the resulting memory layout post-storage. During writing, consecutive elements in a row are distributed across two banks, while consecutive elements in a column are written in the same entry in adjacent banks, enabling the transposed matrix to be retrieved by reading the banks entry-by-entry. In some NoC transactions, element pairs need to be swapped to ensure correct storage order; for instance, elements $(a_{20}, a_{21})$ are swapped before writing, so $a_{20}$ is written into bank 1 and $a_{21}$ into bank 0. Similar swaps occur during reading to ensure accurate transposition during transfer, e.g., when reading $(a_{21}, a_{11})$.

\begin{figure}[t!]
    \centering
    \includegraphics[width=0.9\columnwidth]{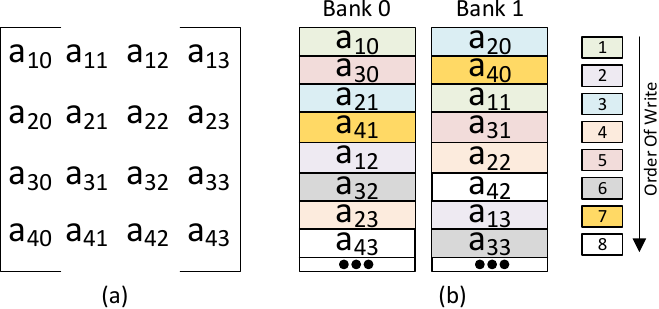}
    \vskip -0.10in
    \caption{(a) Original matrix and (b) Post-storage matrix layout.}
    \vskip -0.25in
    \label{fig:Example_In_Transposition}
\end{figure}

Figure~\ref{fig:Top_view}(d) illustrates our hardware proposal for the transposing bank group, which consists of 16 banks, each with a data width of one byte, and a 16-byte swapping register capable of holding 16 elements of the key matrix. As previously discussed, each attention block in Transformer models includes a FC layer to generate the key matrix. In the Hamun architecture, FC layers operate token-by-token, resulting in a row-by-row generation of the key matrix. Each row is divided into multiple NoC transactions, with each transaction consisting of 16 bytes. These transactions are transferred through the chip's mesh network. Each transaction is further split into 16 distinct elements, and these elements are distributed across the 16 banks in the transposing bank group. The swapping register reorders the elements based on the desired transposed matrix configuration and aligns each elements with its corresponding banks. Equation~\ref{eq:Transposition} (in section~\ref{subs:Transposition}) maps how each element in the original flattened matrix is assigned to its proper position in the transposed matrix. Therefore, to store elements in transposed order, first, the reordering of the elements is executed according to the corresponding bank for each element, as outlined by Equation~\ref{eq:id} below. Next, all elements are simultaneously stored in their respective banks at the entries determined by Equation~\ref{eq:entry}.


\vskip -0.10in
\begin{equation}
    id = P(\alpha) \, mod \, (\#banks)
    \label{eq:id}
\end{equation}

\vskip -0.10in
\begin{equation}
    entry = P(\alpha) \///(\#banks)
    \label{eq:entry}
\end{equation}

The function $P(\alpha)$ defines the correct index in the flattened transposed matrix for an element with index $\alpha$ in the original matrix, as detailed in Equation~\ref{eq:Transposition}. Also, the number of banks, represented by $\#banks$, is set to 16 in the transposing bank group. This in-situ approach eliminates the overhead for transposing the key matrix in Transformer DNNs, as the scheduler statically pre-generates the required swap instructions and bank entry mappings based on the key matrix size and number of banks in the transposing bank group.

\subsection{Execution Dataflow}\label{subs:Execution Dataflow}
n this section, we explain the dataflow of the Hamun accelerator, which involves two distinct procedures: writing weights into the ReRAM crossbars and performing dot-product computations. As previously detailed in Section~\ref{subs:Writing}, the dot products in each crossbar are computed between the input vector and all weights stored in each column of the crossbar. Consequently, when a ReRAM cell becomes worn out, the resulting dot product in the column containing the faulty cell will be incorrect. The most effective solution is to retire the faulty cell, which requires disabling the entire column that includes the damaged cell. In Hamun, this process is handled by masking the faulty column when the MUX in each crossbar connects the column to the ADC, avoiding using the faulty column.



In this paper, a PE row refers to a group of APUs organized in a single row within a PE and share a common buffer for input activations and weights. The PE row serves as the smallest granularity for mapping a DNN layer. Each layer of the DNN is mapped to at least one PE row, where different output neurons (or kernels in convolutional layers) are assigned across distinct APUs within the row. The computations for the same input activation set, across various output neurons, are carried out simultaneously within the PE row. If the number of output neurons exceeds the capacity of a PE row, or if the size of each output neuron surpasses the number of ReRAM cells in an APU column, the layer mapping extends to additional PE rows. This hierarchical organization ensures efficient utilization of resources while accommodating the diverse computational demands of DNN layers.

\emph{Weight Writing Procedure:} As indicated in Figure~\ref{fig:WeightUpdating}, the ReRAM weight writing procedure follows the row-by-row approach within each crossbar. The weight-writing process for a ReRAM crossbar row begins by fetching the weights from main memory. These weights are transferred through the chip NoC to the corresponding PE, where they are stored directly in the destination buffer, bypassing the shift registers. The target APU then reads these weights from the buffer and loads them into its Writing Registers. Simultaneously, mask signals — one bit per column — are fetched from the host and stored in the Mask Register. Note that mask signals are only fetched during the writing of the first row and reused for subsequent rows. The APU drivers are configured to adjust the weights by applying either increasing or decreasing pulses to the ReRAM cells in two phases (Figure~\ref{fig:WeightUpdating}). The SL drivers generate these pulses with varying amplitudes based on the new weights stored in the Writing Registers and the current cell values in the ReRAM according to P\&V scheme. If the mask signal indicates that the column contains a faulty cell, the SL driver is disabled for that column since that column will be masked in computation stage. Meanwhile, the BL driver controls the polarity of the programming signals, as the weight increase and decrease phases require opposite polarities for proper adjustment of the ReRAM cells. The process is repeated for all rows in the crossbar, while simultaneously other crossbars follow the same procedure. Since all cells in a crossbar row are written concurrently, the row's writing latency is governed by the slowest cell in each phase. In other words, the longest latency cell in a row determines the overall write time for that row. Additionally, Hamun takes into account the main memory bandwidth as a limiting factor for the number of APUs can perform weight updates concurrently.
 
In the P\&V scheme, after each programming pulse, the value of each ReRAM cell is read to check whether it has reached the desired state. If a cell fails to change after successive pulses and becomes stuck at a specific value, it is identified as faulty. Whenever a new fault is detected, the accelerator sends a notification to the host, which initiates a routine to isolate the faulty cell that is discussed in more detail in Section~\ref{subs:Scheduler Fault handling}.

\emph{Dot-Product Computations Procedure:} The dot-product computation in Hamun begins by fetching the input activations from on-chip memory (or the main memory in the case of the first layer) and distributing them across the PEs according to the placement of the stored weights. All the weights within a PE row correspond to a single layer and share the same input activation. The activations are then serialized within the PEs, where the PE controller activates the corresponding buffer, and the activations are streamed bit-serially into the buffer. The shift register set, which is responsible to serialize activations, is shared for different PE rows to reduce area overhead. Like other architectures~\cite{ISAAC, PRIME, ReDy}, Hamun iterates over the bits of the activations to perform the dot-product operations (as illustrated in Figure~\ref{fig:CrossBars}(a)). The crossbars compute the partial sums for each activation bit in every iteration, which are then shifted and accumulated across iterations to form the final result. For columns that contain faulty cells, the masking signal is used to disable these columns, preventing incorrect dot-product calculations. The controller, responsible for providing the select signal to the MUX, will skip any column with an active mask signal, ensuring that faulty columns are excluded from the computation process.

In the final iteration, each APU returns its portion of the neural computation, and the complete result for each kernel or filter is computed by aggregating all the APUs' partial sums. Depending on the kernel size and the mapping of the neural network layer, this accumulation can be performed either within the PEs or in the global chip-level accumulator. Once the dot-product calculations are complete, the results are passed through activation, normalization, and pooling functions, after which they are stored in the Global Buffer. The accelerator is then ready to compute the next convolution window in a convolutional layer or process the next token in a Fully Connected layer for tasks like Large Language Models (LLMs).


\section{Hamun Scheduler}\label{s:Scheduler}
Figure~\ref{fig:Compilation} shows the Hamun scheduler, which is divided into multiple main components. The offline scheduler manages both computation and weight-writing tasks by efficiently assigning necessary resources to each task in the Binding procedure, and also orchestrates tasks in the Scheduling procedure. The offline scheduler focuses on two key optimizations aimed at prolonging the accelerator’s lifespan. In contrast, the online component continuously monitors the appearance of worn-out cells and notifies the host of their location. The offline scheduler generates a set of instructions that control both the weight writing procedure and the dot-product computations procedure during each inference. These instructions are then executed by the accelerator for each inference task.

To maintain the \textit{adaptability} of the accelerator as the size of the network grows, Hamun follows a similar execution scheme to those used in previous works~\cite{ARAS, MNEMOSENE}, which emphasize efficient resource utilization. However, Hamun focuses on addressing the primary limitation of these designs, the restricted lifespan of the accelerator due to frequent ReRAM cell updating. Hamun scheduler adopts a systematic layer-by-layer computation strategy and overlaps the computation of the current layer with the writing of weights for subsequent layers. As soon as the computation for a layer is done, the scheduler reallocates the resources used for this computation for writing the weights of subsequent layer(s). Depending on the number of available crossbars, the accelerator can either update the weights for part of a layer, an entire layer, or multiple layers simultaneously.


\begin{figure}[t!]
    \centering
    \includegraphics[width=1.0\columnwidth]{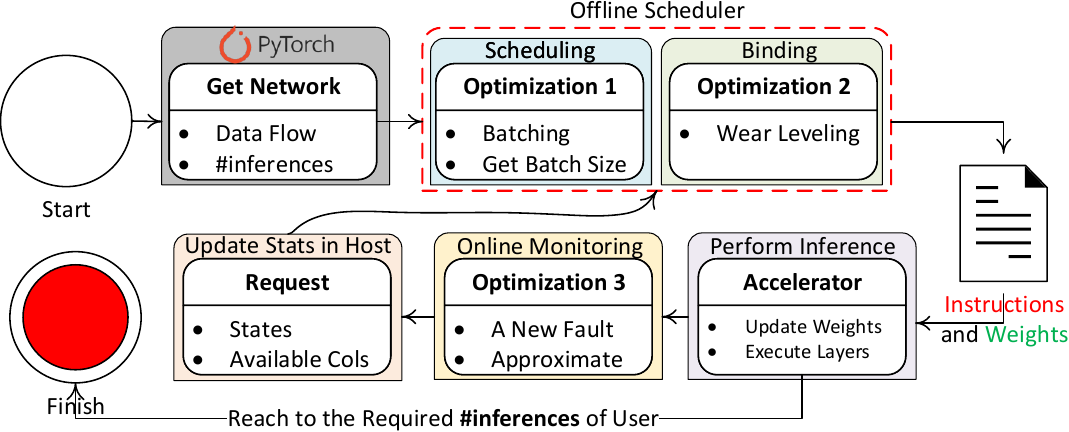}
    \vskip -0.10in
    \caption{Hamun Compilation Flow.}
    \label{fig:Compilation}
    \vskip -0.25in
\end{figure}

Figure~\ref{fig:Execution_Flow} shows an example of the Hamun execution flow for an encoder block commonly used in LLMs. Initially, based on the available accelerator ReRAM crossbar resources, the weights for multiple layers are written into the ReRAM crossbars. In this example, the first three FC layers responsible for generating the Query, Key, and Value matrices, along with the final FC layer in the attention block, are written simultaneously. Hamun performs matrix multiplication using ReRAM crossbars, where one matrix play the role of weights and the other play the role of activations. Once the Query, Key, and Value matrices are produced, the resources used for their respective FC layers are released, and the Key and Value matrices are written into ReRAM memory to perform the matrix multiplication required to get the score matrix. It is important to note that before writing, the Key matrix must be transposed. Hamun utilizes in-situ transposition for this purpose, as described in section~\ref{subs:Architecture}. Once the matrix multiplications are complete, the assigned ReRAM crossbars are released, allowing the weights for the final two feed-forward FC layers to be written. As illustrated, Hamun follows a layer-by-layer computation approach, respecting the data dependencies between layers to ensure correct execution. Moreover, Hamun overlaps the weight writing process with the computation of dot products to effectively hide the latency associated with costly ReRAM writes, optimizing both time and resource utilization.

The key insight in Hamun's offline scheduling and binding process lies in its dual strategy of overlapping layer computations with the writing of weights for the subsequent layers, alongside efficient resource management to maximize parallelism in the weight update process. By executing the computation of one layer while concurrently writing the weights for the next layer(s), Hamun reduces latency and ensures that weight writing doesn't become a bottleneck. Moreover, efficient resource allocation allows multiple weights to be updated simultaneously, which not only increases performance but also ensures faster transitions between layers.

\begin{figure}[t!]
    \centering
    \includegraphics[width=1.0\columnwidth]{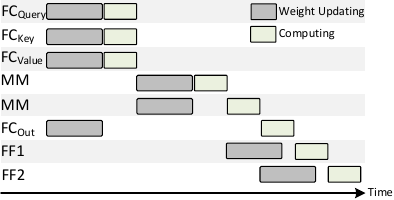}
    \vskip -0.10in
    \caption{An example of the execution flow in Hamun for an encoder block typically found in LLMs, showcasing its layer-by-layer computation scheme, which efficiently overlaps weight updates with ongoing computations.}
    \label{fig:Execution_Flow}
    \vskip -0.25in
\end{figure}

\subsection{Hamun Fault Handling}\label{subs:Scheduler Fault handling}
In this section, we describe the wear-out fault-handling mechanism employed by Hamun. As mentioned earlier, the Program \& Verify (P\&V) scheme used for updating ReRAM cells can detect cells that become stuck at a specific value (permanent failure) and do not respond to programming pulses. If a ReRAM cell fails due to reaching its endurance limit while being updated, the accelerator immediately alerts the host system via the online monitoring component (Figure~\ref{fig:Compilation}). Hamun's fault-handling routine estimates the impact of this faulty cell in the final accuracy, and if it is above the user-specified threshold it retires the column where the faulty cell resides and re-does the scheduling and binding process to ensure accurate DNN inference despite the fault.

Figure~\ref{fig:Compilation} illustrates the Hamun compilation process when a new cell wears out during inference execution. The process starts with analyzing the PyTorch model to extract the network structure and data dependencies. Following this, the scheduling and binding tasks are performed offline, based on the network structure and its dependencies. Two key optimizations are applied during this phase. The first is network batch execution, where weights are reused across different inferences within a batch, reducing the overhead of multiple memory writing operations for each inference. The second optimization focuses on resource management during the binding process, where resources are prioritized according to their wear level. The wear level is tracked using counters that monitor the frequency of resource usage, this data being stored in the host side to avoid adding extra area overhead to the accelerator. To ensure balanced writing distribution across the ReRAM crossbars, those crossbars that have been utilized less frequently are given higher priority in the binding process to assign new computations to them. This approach optimizes resource usage and prolongs the lifespan of the ReRAM cells by evenly distributing the wear.

After the offline scheduler generates the instructions for executing network inferences, the accelerator proceeds with performing inferences based on this configuration. When a new ReRAM cell wears out, the accelerator notifies the host of the faulty cell's location and the current inference execution stops. At this point, during the "Request" stage, the wear levels of resources are updated based on the number of inferences executed with the current scheduling and binding configuration. A new request is then sent to the scheduler, prompting it to redo the scheduling and binding processes to generate a new configuration that may exclude the worn-out cell while maintaining full network accuracy.


Hamun introduces an approximation optimization that, instead of retiring a faulty cell immediately after its detection, it determines its impact on accuracy together with previously detected and not retired cells. If the collective impact do not significantly degrade accuracy, the current scheduling and binding configuration is maintained, allowing continued operation. However, if the expected accuracy loss exceeds a user-defined acceptable threshold, all faulty cells are retired simultaneously, and a new configuration is computed. This strategy ensures minimal interruptions and maximizes the accelerator's operational lifespan by only retiring cells when absolutely necessary.

When a ReRAM cell becomes worn out, the entire column containing the faulty cell in the crossbar is deactivated by disabling the corresponding mask signal to prevent incorrect dot-product computations. Consequently, during the re-binding process, no weights from any layer are assigned to that specific column. This ensures that the faulty column is not used in future computations. Multiple PE rows are utilized for a layer when the size of the input exceeds the number of cells available in a crossbar column. In such cases, since the final result of the dot-product is obtained by accumulating partial results from all APUs within each PE column, it is crucial that the number of crossbar columns to which the layer's weights are mapped remains consistent across all APUs in a PE column. Therefore, the number of crossbar columns used in each APU is aligned with the minimum number of available columns across all crossbars within a PE column, providing balanced resource utilization and accurate accumulation of results.

At first glance, aligning the number of APUs' columns within each PE column may appear to underutilize the accelerator's resources. However, if the number of columns used in a crossbar is lower than the total number of available columns on that crossbar, this alignment introduces an advantage in the event of a cell in the used columns becoming faulty. In such scenarios, the scheduler can reassign one of the unused columns within the crossbar to replace the faulty one.

This process repeats until the accelerator executes the specified number of inferences set by the user. Throughout the operation, the system continually monitors the health of the ReRAM cells and dynamically updates the scheduler whenever a new worn-out cell is detected.

\subsection{Wear Leveling Techniques}\label{subs:WL}
As illustrated in Figure~\ref{fig:Top_view}, the Hamun system architecture is organized into multiple levels, enabling the implementation of Wear Leveling (WL) techniques in different levels of abstraction. Each PE row includes a counter that tracks the frequency of its use, referred to as the "wear level" of that row. These counters are stored on the host side, and during the binding process, the offline scheduler uses these wear levels to efficiently assign resources for each layer. These counters are updated at "Request" stage according to number of inferences that are executed with the current binding and scheduling configuration.

Hamun employs a counter per PE row rather than per ReRAM crossbar row or per cell for two main reasons. First, having a counter for each individual crossbar cell or row would significantly increase the complexity of the scheduling and binding process, thereby slowing down the offline scheduler's execution time. Managing counters at such a fine-grained level would introduce substantial overhead in updating each cell's usage, especially as faults emerge. Second, Hamun includes a crossbar-level WL technique that eliminates the need to monitor the frequency of cell or row utilization within each crossbar. This approach ensures balanced wear across cells without adding per-cell tracking, streamlining both system efficiency and durability.

The wear level of each PE is determined by the maximum wear level among all of its rows. During the binding process, the scheduler first sorts the available PEs based on their wear level index. Then, for each layer, the scheduler selects the PEs with the lowest wear levels to evenly distribute wear across the system. Similarly, when assigning rows within a PE, Hamun also applies WL techniques, prioritizing rows with lower wear level counters.

As mentioned before, Hamun also employs WL techniques at the crossbar level to enhance the longevity of ReRAM cells. During the weight-writing process in the ReRAM crossbar, cells storing the least significant bits (LSBs) experience a higher frequency of updates compared to those storing the most significant bits (MSBs). To address this imbalance, Hamun utilizes a byte-granularity WL scheme. For each inference, this scheme remaps bit positions to different ReRAM cells in a round robin manner. Figure~\ref{fig:wl}(a) provides an example, in which each cell stores two bits, meaning that four cells are required to store an 8-bit weight. In this example, the two LSB bits of an 8-bit weight are initially stored in \textit{cell 0} of a given set of cells. Then, the WL technique reassigns these bits to \textit{cell 3} in the subsequent inference, so update frequency is distributed evenly across all cells. In addition, this remapping is seamlessly integrated with the dot-product computation by aligning the ADC column sampling order with the remapped bit positions.

Figure~\ref{fig:wl}(b) illustrates another wear leveling (WL) technique employed by Hamun, designed to evenly distribute wear across ReRAM crossbar rows. When a layer’s weights do not fully occupy all available rows in a crossbar, utilizing the same set of physical address for each inference leads to uneven wear in specific rows. This repeated ReRAM update causes certain rows to undergo more frequent write cycles, leading to premature degradation of those crossbar rows while underutilizing others. To mitigate this, Hamun shifts the starting point of the weight-writing procedure to a different physical address for each inference. As shown in Figure~\ref{fig:wl}(b), for each inference, the starting row of the weight-writing process is incremented by one. This dynamic shifting ensures that all rows wears out at the same pace, extending the life of the ReRAM crossbar. To ensure accuracy in the dot-product computation, the order of activations fetched into the APU’s input register is adapted so that each activation is correctly multiplied by its corresponding weight.

\begin{figure}[t!]
    \centering
    \includegraphics[width=1.0\columnwidth]{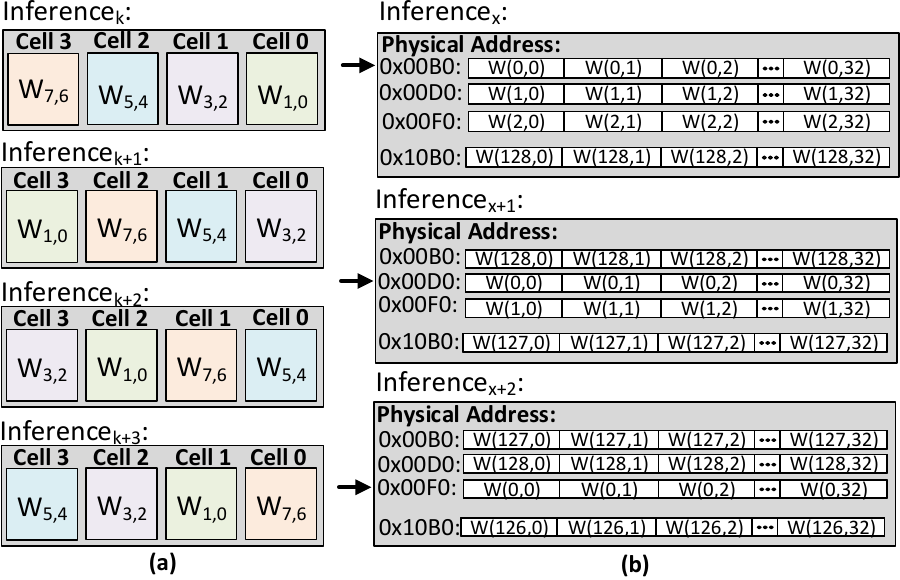}
    \vskip -0.1in
    \caption{Wear leveling techniques at the crossbar abstraction level. (a): First scheme remaps bit positions to different ReRAM cells for each inference. (b): Second scheme shifts the starting row of the weight-writing procedure in each crossbar for each inference.}
    \label{fig:wl}
    \vskip -0.20in
\end{figure}

The proposed WL schemes in Hamun aim to evenly distribute write operations across all cells in a ReRAM crossbar. Another related approach, proposed by ARAS~\cite{ARAS}, takes a different route by increasing the similarity across the weights of different layers during inference, which reduces the number of weight updates required. While this method successfully decreases the total weight update frequency, it leaves the "hotspot" issue unresolved, especially in cells storing the LSBs, which still experience most of the updates. However, when Hamun's WL techniques are applied on top of ARAS's scheme, the reduced number of total weight updates results in even greater improvements in the accelerator's lifespan. This is because Hamun's WL scheme ensures that the reduced number of weight updates is more evenly distributed across all cells, addressing the imbalance between LSB and MSB cells, and mitigating the wear hotspot issue. Consequently, the combination of both approaches maximizes the lifespan of ReRAM cells by reducing and evenly distributing the frequency of updates.

\subsection{Batch Execution}\label{subs:Batch}
The weight updating procedure in ReRAM-based accelerators is a costly process in terms of energy consumption, latency, and its impact on the lifespan of memory cells. To mitigate this cost, Hamun employs a batch execution technique. This method allows the accelerator to execute multiple inferences without reallocating resources or rewriting weights between each inference. Essentially, the weights for a given layer are written into the ReRAM crossbars once, and the same set of weights is reused for consecutive inferences. The scheduler in Hamun still follows the same execution flow as described above, but instead of immediately reallocating resources to the next layer after completing one, it processes multiple inferences using the current layer's weights. While Hamun’s batch execution technique helps reduce the cost of frequent weight updates, it does generate more partial results during the computation process, which increases pressure on the on-chip SRAM memory.

\begin{figure}[t!]
    \centering
    \includegraphics[width=0.95\columnwidth]{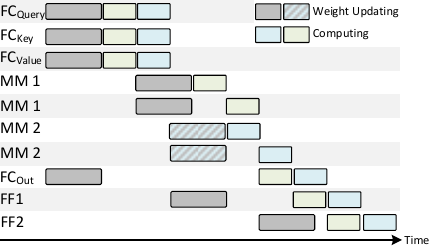}
    \vskip -0.10in
    \caption{Batch Execution with size two. Blue blocks represent computation steps for a new inference, while patterned blue and gray blocks indicate additional writes required by the second inference.}
    \label{fig:Batching}
    \vskip -0.25in
\end{figure}

Figure~\ref{fig:Batching} illustrates the batch execution of the previous example (Figure~\ref{fig:Execution_Flow}). For layers with static weights, such as Fully Connected and Convolutional layers, batch execution allows the reuse of written weights across multiple inferences, with the corresponding partial results stored separately in on-chip memory. This means that once a set of weights is written, they are reused for all computations associated with multiple inferences before being overwritten by new weights. Even in cases where the accelerator only has enough ReRAM storage to accommodate a portion of a layer, all computations for that portion across different inferences are completed before reallocating the ReRAM for the next portion.

In cases where a layer does not have static, pre-known weights such as matrix multiplication in an attention block, batch execution becomes more complex. Specifically, executing that layer for multiple inferences requires a separate weight update for each inference, as one of the operands in the matrix multiplication must be written to ReRAM for every inference (shown in patterned blue and gray color in the figure). This weight updating procedure begins as soon as the operand is computed by the preceding layer. Unlike layers with static weights, where Hamun achieves significant lifespan improvements by reusing the same weights across multiple inferences, non-static layers do not benefit from this optimization.

The number of inferences that can be batched together in Hamun depends on the available on-chip SRAM memory. A larger batch size leads to more inferences being processed simultaneously, which generates more partial results, requiring additional on-chip memory for storage. To optimize this, Hamun employs an offline iterative procedure to determine the maximum number of DNN inferences that can be included in a batch. In each iteration, the procedure increments the batch size and calculates the required on-chip memory. The process continues until the memory requirement exceeds the accelerator's on-chip memory capacity. Once the optimal batch size is identified, Hamun sets this value for its offline scheduler, which then generates the appropriate instructions for executing the batch. This approach maximizes the number of inferences processed in parallel, increasing the reuse of written weights in the ReRAM crossbar. As a result, it significantly improves the lifespan of the accelerator by allowing it to execute a greater number of DNN inferences.

\subsection{Approximate Computing}\label{subs:Approximation}
The inherent fault tolerance of deep neural networks (DNNs) allows them to accommodate some inaccuracies in their computations without significantly affecting performance. This resilience, stemming from the redundant and distributed nature of DNN learning processes, enables DNNs to tolerate occasional weight errors. Such fault tolerance is particularly advantageous in ReRAM-based accelerators, where cell wear-related inaccuracies can often be tolerated without requiring immediate retirement of faulty cells. This tolerance can extend the accelerator’s lifespan by reducing the need for frequent reconfiguration.


As detailed in Section~\ref{subs:Scheduler Fault handling}, when a ReRAM cell wears out, the scheduler notifies the host via the online monitoring component. Depending on the DNN layer size and binding configuration, a worn-out cell can impact one or multiple layers or even introduce multiple faults within a single layer. To extend the accelerator’s lifespan, Hamun introduces a fault-tolerant strategy that optimizes fault handling. Rather than retiring each faulty cell immediately, Hamun assesses the fault’s impact on inference accuracy. If the accuracy loss is greater than a user-defined threshold, the faulty cell is retired, and a new scheduling and binding configuration is computed. Otherwise, the retirement is deferred, allowing the accelerator to continue with the existing configuration.

As additional cells wear out, Hamun repeats this process by evaluating the cumulative effect of these multiple faults. This iterative assessment continues until the accumulated faults degrade accuracy beyond the acceptable threshold, at which point all faulty cells are retired, and the system is reconfigured to maintain accuracy. This approach minimizes interruptions by retiring cells only when absolutely necessary, thereby maximizing the operational lifespan of the accelerator.

To gauge the impact of faulty cells on accuracy, Hamun employs an offline estimation method to ensure that accuracy loss remains within the acceptable threshold. During this process, Hamun randomly introduces an equal number of faults across all layers and evaluates the inference accuracy using the entire validation or test dataset. These faults are applied to random weights and random bit positions within those weights, ensuring a general assessment of network robustness. Moreover, to evaluate diverse fault scenarios, for each inference in the validation or test dataset, a different set of random faults is imposed.

Through a iterative process, Hamun progressively increases the fault count uniformly across all layers until accuracy degradation exceeds the user-defined threshold. This provides an \textbf{optimal fault tolerance threshold}, representing the network’s resilience to inaccuracies. This threshold is used at runtime to guide decisions on whether to postpone retirement of faulty cells without compromising network accuracy.

At runtime, when a fault occurs, Hamun’s host system identifies the affected layer(s) based on the fault’s location. To evaluate potential accuracy loss from multiple faults, Hamun tracks the number of weights impacted within each layer. If this number surpasses the \textbf{fault tolerance threshold} for any layer, Hamun flags the accuracy impact as unacceptable. In such cases, all faulty cells are retired, and the system redoes the binding and scheduling to maintain inference accuracy. This method ensures that accuracy loss remains within the user-specified limits, optimizing both performance and lifespan.


\section{Methodology}\label{s:Methodology}
We developed an event-driven simulator to accurately model the lifespan of Hamun, and compare it against ARAS~\cite{ARAS}, which serves as our baseline. Lifespan is defined as the number of inferences a ReRAM-based accelerator can perform while maintaining an acceptable level of throughput. In ARAS, the simulation ends as soon as any ReRAM cell reaches a wear-out point, as it lacks a fault-handling mechanism to deal with faulty cells. In contrast, Hamun is designed with a fault-tolerant approach that allows it to continue functioning even when individual cells wear out. It maintains operation until throughput degradation reaches a user-defined threshold, allowing for greater longevity and resilience in performance. For this simulation, we set the acceptable throughput drop to $40\%$ of the maximum accelerator throughput.

The evaluation of area, latency, and energy consumption for the proposed ReRAM-based accelerator leverages a multi-tool methodology for detailed component modeling. ReRAM crossbars are simulated using NeuroSim~\cite{Neurosim_github}. For on-chip buffers, CACTI-P~\cite{cacti-p} is used, and logic components, such as control and computation units, are implemented in Verilog and synthesized using Synopsys Design Compiler~\cite{Design_compiler} with a 28/32nm technology library. Main memory is assumed to be an LPDDR4 module with 8 GB capacity and 19.2 GB/s bandwidth (single-channel) and is simulated using DRAMSim3~\cite{DRAMsim3}.

To evaluate the accelerator's lifespan, each ReRAM cell is initialized with a specific endurance (number of writes before wear-out). The initial endurance of the cells follow a normal distribution, as established in \cite{Realizing}, with a Coefficient of Variation (CoV) of $0.2$ and a mean value of $2.5\times10^9$.

For a fair comparison with the ARAS baseline, we configured the accelerator with similar parameters, as outlined in Table~\ref{tab:Param}. Specifically, we set the accelerator to include 64 PEs, each composed of 6x4 APUs. Each APU features a crossbar array of $128\times128$ ReRAM cells, and each ReRAM cell has a 2-bit storage resolution. Consequently, representing an 8-bit weight requires four consecutive ReRAM cells. Furthermore, the batch size, or the number of inferences processed concurrently, depends on the available on-chip SRAM memory as explained above. To consider adequate on-chip memory capacity and enable fair comparison, we configure this parameter at 8 MB, in line with the Google Edge TPU’s on-chip SRAM size.

\begin{table}[t!]
\caption{Hamun accelerator configuration parameters.}
\label{tab:Param}
\centering
\resizebox{0.6\columnwidth}{!}{%
    \centering
    \begin{tabular}{|c|c|}
    \hline
    \cellcolor[gray]{0.9} Technology & 32 nm \\
    \cellcolor[gray]{0.9} Frequency & 1 GHz \\ 
    \cellcolor[gray]{0.9} Number of ADCs per APUs & 16 \\
    \cellcolor[gray]{0.9} ADCs Sampling Precision & 6-bits \\
    \cellcolor[gray]{0.9} PE Buffers Size & 1.5 KB \\
    \cellcolor[gray]{0.9} Crossbar Computation Latency & 96 Cycles \\
    \cellcolor[gray]{0.9} Crossbar Row Writing Latency & 6000 Cycles \\
    \hline
    \end{tabular}%
}
\vskip -0.20in
\end{table}

We evaluate our scheme on three prominent DNNs with distinct architectures and applications: Vision Transformer (ViT)~\cite{ViT} for image classification, BERT~\cite{BERT} for question-answering, and GPT-2~\cite{GPT2} for text classification. The Vision Transformer is assessed on the ImageNet~\cite{ImageNet} dataset, which is widely used in image classification tasks. BERT is evaluated on the SQuAD v1.1~\cite{SQuAD} dataset, a benchmark for question-answering tasks that requires the model to identify answers within a context passage. GPT-2, a decoder-based language model, is tested on the IMDB~\cite{imdb} dataset for text classification, specifically for sentiment analysis. Both BERT and ViT contain 12 encoder blocks, while GPT-2 is composed of 12 decoder blocks. The INT8 accuracy (F1) for GPT-2 and BERT are $91.37\%$ and $79.73\%$ respectively, while the accuracy (Top-1) for ViT is $80.96\%$. In the Hamun approximation scheme, the maximum acceptable accuracy loss is set at $1\%$, although this threshold can be adjusted according to user requirements.

\section{Experimental Results}\label{s:Results}
This section evaluates the performance degradation of the Hamun accelerator over time, its lifespan improvement, and the scheduler reconfiguration overhead. First, we present the progression of retired crossbar columns over time and assess how this impacts overall performance degradation. Second, we analyze the overall lifespan improvements achieved by Hamun, breaking down the contributions of individual optimizations to total accelerator longevitylly, we analyze the offline scheduler overhead, highlighting the cost introduced by the re-scheduling task in response to new faults. This analysis also examines each optimization’s contribution to reducing reconfiguration overhead.

\subsection{Hamun Performance Evaluation}\label{subs:trend}
Figure~\ref{fig:Performance} illustrates the Hamun gradual performance decline over time as ReRAM cells undergo wear and experience retirement. As noted in Section~\ref{s:Methodology}, Hamun is designed to operate until performance degradation reaches a 40\% throughput drop, at this point it halts operations. However, this threshold is adaptable based on user requirements. For instance, in applications where real-time throughput is acceptable, the performance threshold could be set to a lower level to ensure consistent real-time processing and prolong the lifespan of the accelerator.

As shown in the figure, the accelerator maintains peak performance for a significant portion of its operational lifespan. This demonstrates the effectiveness of Hamun's optimizations, such as fault-handling and wear-leveling, in sustaining throughput and minimizing performance drops. The lifespan data, expressed in days, is based on a utilization rate assumption of 25\%. Across all benchmarks, the lifespan of ReRAM-based accelerators remains limited to about a year due to ReRAM cell endurance constraints, even with the significant lifespan improvements achieved by Hamun. This restricted longevity highlights the necessity of advancements in ReRAM technology to establish ReRAM-based accelerators as competitive options against other computation platforms.

Hamun does not change the number of ReRAM cells that require writing in the accelerator compared to the baseline, which is determined by the DNN model size. Consequently, the energy consumption for weight writing, the largest contributor to energy usage in each inference (as reported in \cite{ARAS}), remains unchanged from the baseline. Since the accelerator’s throughput diminishes over time, the static energy per inference, which contributes slightly in total energy, increases gradually and leads to a small increase in total energy consumption.


\begin{figure}
    \centering
    \begin{subfigure}{\columnwidth}
        \centering
        \includegraphics[width=0.8\textwidth]{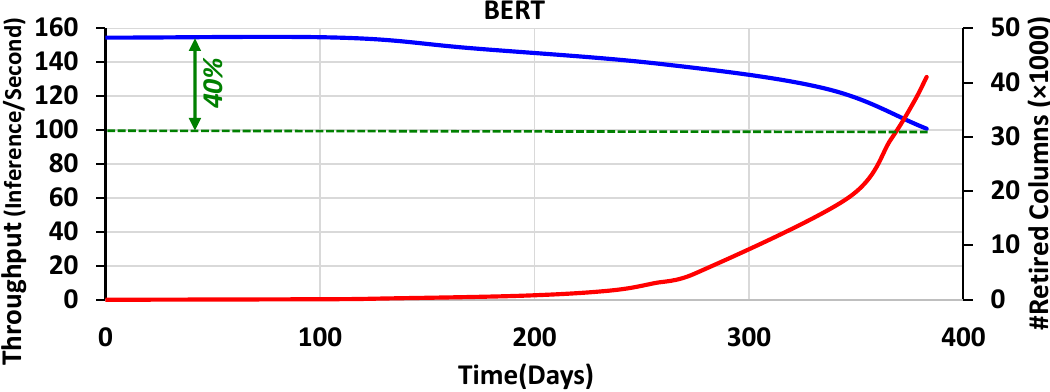}
        \label{fig:first}
    \end{subfigure}
    \vskip +0.10in
    \begin{subfigure}{\columnwidth}
        \centering
        \includegraphics[width=0.8\textwidth]{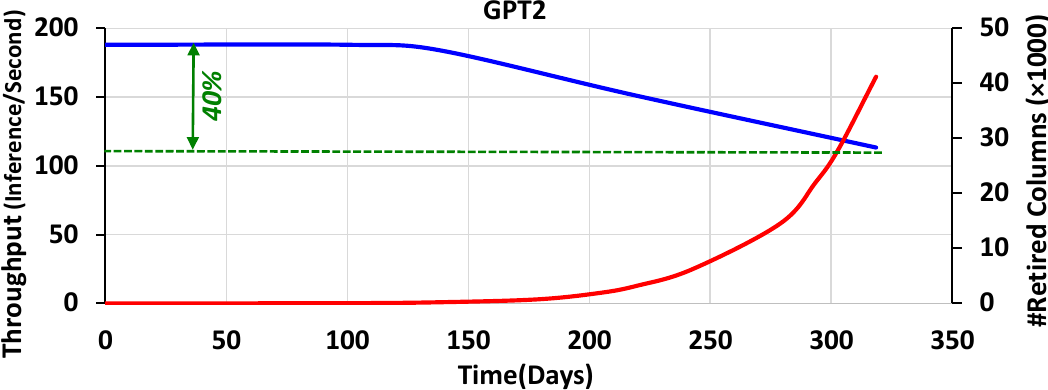}
        \label{fig:second}
    \end{subfigure}
    \vskip +0.10in
    \begin{subfigure}{\columnwidth}
        \centering
        \includegraphics[width=0.8\textwidth]{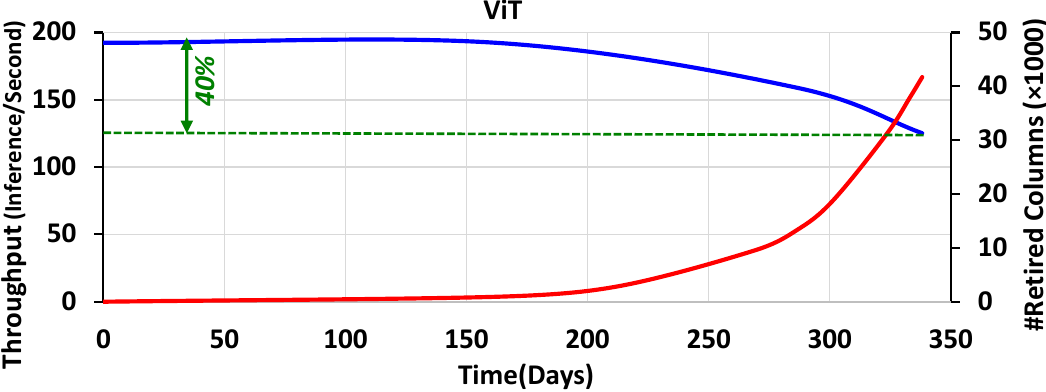}
        \label{fig:third}
    \end{subfigure}
    \caption{Performance degradation and number of retired columns over time.}
    \vskip -0.10in
    \label{fig:Performance}
\end{figure}

\subsection{Hamun Lifespan Improvement}\label{subs:Lifespan}
Figure~\ref{fig:lifespan_Improvement} illustrates the lifespan improvements achieved by Hamun over the baseline, with or without the fault-handling mechanism. Additionally, the figure breaks down the contributions of each optimization to the total lifespan enhancement. As explained in Section~\ref{s:Methodology}, the baseline ceases operation once any ReRAM cell fails. In contrast, Hamun employs a fault-handling mechanism that retires faulty cells and dynamically generates new scheduling and binding configurations, allowing the accelerator to continue functioning until the performance degradation reaches the user-defined threshold.


Among the optimizations, the batch execution technique and fault-handling mechanism have the largest impacts on extending lifespan. Overall, Hamun delivers a substantial improvement, extending ReRAM-based accelerator lifespan by $13.2\times$ over the baseline. Specifically, $4.6\times$ improvement is due to fault handling, while batching adds another $2.6\times$. Both wear-leveling and approximate computing provide an additional $1.1\times$ improvement, which although not negligible their contributions are overshadowed by the huge impact of fault handling. Notably, in configurations where fault handling is not included, both wear leveling and approximation significantly enhance lifespan by $2.5\times$, underscoring their effectiveness in isolation. Moreover, wear-leveling and approximate computing significantly reduce the overhead of reconfigurations, a topic discussed in more detail in the following section.

\begin{figure}[t!]
    \centering
    \includegraphics[width=1.0\columnwidth]{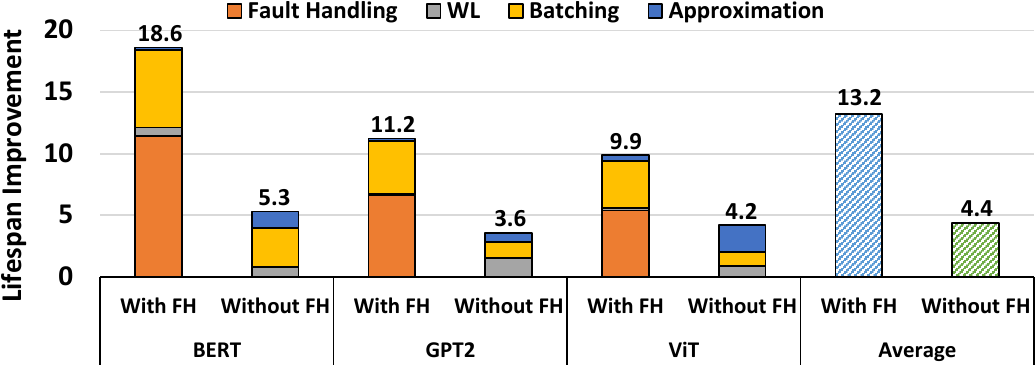}
    \vskip -0.05in
    \caption{Breakdown of each optimization’s contribution to lifespan enhancement over the baseline (ARAS).}
    \label{fig:lifespan_Improvement}
    \vskip -0.20in
\end{figure}

\subsection{Hamun Reconfiguration Overhead}\label{subs:Scheduler_Utilizatio}
Figure~\ref{fig:Reconfiguration} presents the average number of inference executions that can be processed between two consecutive reconfiguration operations for different Hamun configurations. Hamun optimizations reduce the frequency of reconfigurations, which require to suspend inference operations on the accelerator and consume energy in the host system to generate new scheduling and binding configurations. By increasing the number of inference executions per reconfiguration, Hamun reduces reconfiguration overhead, thereby optimizing both performance and energy efficiency.

As Figure~\ref{fig:Reconfiguration} shows, the fault handling optimization alone, when executing the ViT model, requires a reconfiguration after approximately $2.2 \times 10^{5}$ inferences. When wear-leveling (WL) is applied in addition to fault handling, the number of inference executions per configuration increases to around $3.1 \times 10^{5}$. Unlike the lifespan improvement, the approximation scheme has the most significant impact on extending configuration reuse and thus, reducing reconfiguration overhead, allowing ViT to reach $4.18 \times 10^{6}$ inference executions before requiring reconfiguration. This demonstrates the effectiveness of Hamun in reducing the frequency of re-scheduling, thus minimizing performance overhead associated with reconfiguration. With the configuration defined for the accelerator size described in Section~\ref{s:Methodology}, Hamun supports approximately 2.9 million inferences on average across the benchmark set without requiring reconfiguration interruptions.

\begin{figure}[t!]
    \centering
    \includegraphics[width=1.0\columnwidth]{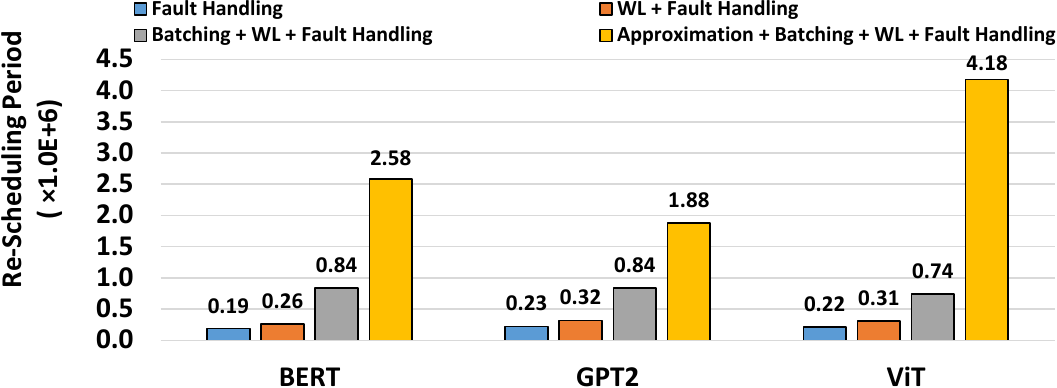}
    \vskip -0.05in
    \caption{Average number of inferences per reconfiguration.}
    \label{fig:Reconfiguration}
    \vskip -0.20in
\end{figure}

\section{Conclusions}\label{s:Conclusion}
In this paper, we demonstrate that frequent ReRAM cell updates needed for DNN inference significantly shorten the lifespan of ReRAM-based accelerators due to the limited endurance cycles of ReRAM cells. To address this challenge, we introduce \textit{Hamun}, an approximate computation method designed to extend the lifespan of ReRAM-based accelerators through multiple optimizations. Hamun features a novel fault-handling scheme that identifies worn-out cells and retire them to prevent their impact on DNN accuracy. Additionally, Hamun employs wear-leveling and batch execution techniques to further increase longevity. To reduce the overhead of retiring cells, Hamun also incorporates an approximation method, thereby extending the accelerator’s lifespan with minimal degradation in DNN accuracy. Across a set of popular DNNs, Hamun achieves a $13.2\times$ lifespan improvement over the baseline on average, highlighting its potential in making ReRAM-based accelerators more viable for long-term use.

\section{Acknowledgments}\label{s:Acknowledgments}

This work has been supported by the CoCoUnit ERC Advanced Grant of the EU’s Horizon 2020 program (grant No 833057), the Spanish State Research Agency (MCIN/AEI) under grant PID2020-113172RB-I00, the Catalan Agency for University and Research (AGAUR) under grant 2021SGR00383, and the ICREA Academia program. 
}

\bibliographystyle{ACM-Reference-Format}
\bibliography{refs}

\end{document}